\documentclass[12pt]{article} 
\usepackage{graphicx}
\usepackage{color}
\definecolor{lgrey}{rgb}{0.3,0.3,0.3}
\definecolor{dgrey}{rgb}{0.6,0.6,0.6}
\definecolor{lblue}{rgb}{0.4,0.4,1.0}
\definecolor{dblue}{rgb}{0.0,0.0,0.6}
\definecolor{dgblue}{rgb}{0.2,0.4,0.4}
\definecolor{lred}{rgb}{1.0,0.4,0.4}
\definecolor{dred}{rgb}{0.65,0.0,0.0}
\definecolor{lgreen}{rgb}{0.4,1.0,0.4}
\definecolor{dgreen}{rgb}{0.0,0.4,0.0}
\definecolor{lbrown}{rgb}{0.45,0.3,0.3}
\definecolor{brown}{rgb}{0.55,0.45,0.45}
\definecolor{dbrown}{rgb}{0.75,0.6,0.6}
\definecolor{lpurple}{rgb}{0.8,0.4,1.0}
\definecolor{purple}{rgb}{0.4,0.0,0.6}
\definecolor{dpurple}{rgb}{0.5,0.0,0.7}

\newcommand{\er}{$\epsilon_{rec}$}

\newcommand{\EE}[1]{\times 10^{#1}}
\newcommand{\elp}{e^{+}}
\newcommand{\elm}{e^{-}}
\newcommand{\mup}{\mu^{+}}
\newcommand{\mum}{\mu^{-}}
\newcommand{\pip}{\pi^{+}}
\newcommand{\pim}{\pi^{-}}
\newcommand{\nue}{\nu_{e}}
\newcommand{\num}{\nu_{\mu}}
\newcommand{\anue}{\bar{\nu}_{e}}
\newcommand{\anum}{\bar{\nu}_{\mu}}
\newcommand{\ceren}{${\rm \check{C}}$erenkov }

\begin{document} 

\begin{titlepage} 
\rightline{FERMILAB-CONF-04-196-E}
\rightline{\today}
\vspace*{4cm}
\begin{center}
{\Large {\bf Low-Energy Neutrino Beams with an} \\[0.1cm]
        {\bf Upgraded Fermilab Proton Driver}} \\
\vspace*{0.5cm}
{\bf S.J.~Brice, S.~Geer} \\
Fermilab \\[0.25cm]
{\bf K.~Paul} \\
University of Illinois \\[0.25cm]
{\bf R.~Tayloe} \\
Indiana University
\vspace*{1cm}
\begin{abstract}
The beam properties of low energy neutrino beams that would be possible
at an upgraded 2 MW Fermilab 8 GeV Proton Source are described. In particular
three options are considered; (i) a MiniBooNE-like beam using a
conventional neutrino horn, (ii) a decay at rest neutrino source, and
(iii) a neutrino beam downstream of a solenoid decay channel. In all
three cases the fluxes are sufficiently large to provide an interesting
physics program. Some physics examples are considered.
\end{abstract}
\end{center}
\vfill
\end{titlepage} 

\begin{section}{Introduction}
\label{sect:intro}
The Fermilab Long Range Planning Committee \cite{LRPReport:2004} has considered 
which facilities and associated physics programs might provide an exciting 
future for the laboratory beyond the Tevatron Collider era. Two 
accelerator-facility-based physics programs have emerged as leading candidates 
for Fermilab's future: (i) a Linear Collider and (ii) an expanded neutrino 
oscillation physics program driven by a new high intensity 8 GeV  proton 
source delivering a 2 MW beam on target \cite{PDriverStudy:2002}. A small 
fraction of the beam from the new proton source would be injected into the 
Fermilab MI to provide a higher energy (up to 120 GeV) proton beam, also with 
2 MW beam power on target. This scheme would enable the intensity of the NuMI 
neutrino beam to be increased by a factor of about 5 for on-axis and/or 
off-axis long-baseline experiments, whilst at the same time providing an almost 
2 MW proton beam at 8 GeV for lower energy neutrino experiments. In this note 
we describe the properties of various low energy neutrino beams that could be 
created with a 2 MW proton beam at 8 GeV.
\end{section}

\begin{section}{A High Intensity MiniBooNE-Like Beam}
\label{sect:MiniBooNE}


\begin{figure}
\begin{center}
\includegraphics[width=3.5in]{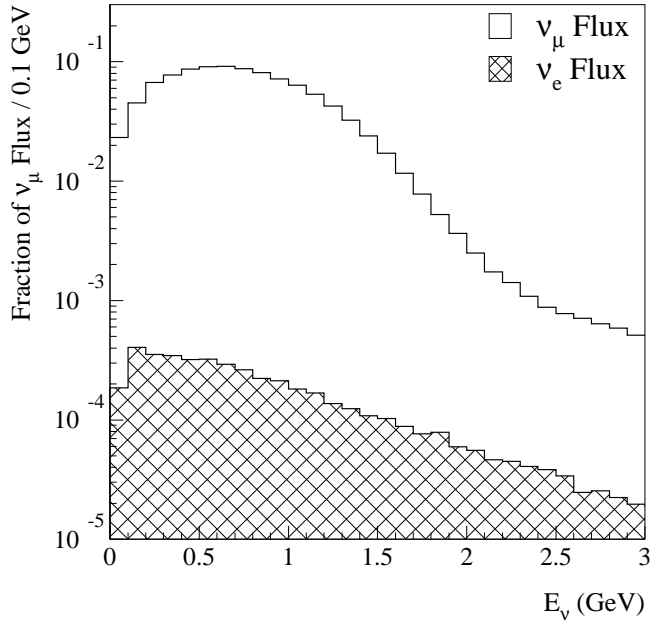}
\caption{The simulated $\num$ and $\nue$ flux energy spectra at the MiniBooNE 
detector. (From \cite{MBRunPlan:2003}).}
\label{fig:miniboone_nuspec}
\end{center}
\end{figure}

The existing MiniBooNE beamline and detector \cite{MBRunPlan:2003} provide a 
useful point of comparison for neutrino beams at a future proton driver. 
MiniBooNE takes 8 GeV protons from the FNAL Booster, steers them into a 
Beryllium target, and uses a magnetic horn to focus the resulting mesons. The 
mesons are given 50m to decay before hitting a concrete absorber. The resulting 
neutrino beam crosses 500m of dirt before reaching the MiniBooNE detector, a 
6.1m radius spherical tank of mineral oil viewed by 1520 PMTs, looking at the 
\ceren and scintillation light created when neutrinos interact and produce 
charged particles in the tank. Fig.~\ref{fig:miniboone_nuspec} shows the 
simulated $\num$ and $\nue$ energy spectra at the MiniBooNE detector. 

\begin{table}
\begin{center}
\begin{tabular}{|c|c||} \hline
                          &                   \\[-0.5cm]
                          &  Million events   \\
$\nu_\mu$ event type      &     per year      \\ \hline
                          &                   \\[-0.25cm]
$\nu_\mu n\rightarrow\mu^-p$ (CC QE)        & $7.7$ \\[0.25cm]
$\nu_\mu\; (n/p)\rightarrow\nu_\mu\; (n/p)$ (NC QE)        & $3.2$ \\[0.25cm]
$\nu_\mu\; (n/p)\rightarrow\mu^-\pi^+\;(n/p)$ (CC $\pi^+$)    & $4.5$ \\[0.25cm]
$\nu_\mu n\rightarrow\mu^-\pi^0p$ (CC $\pi^0$)    & $0.8$ \\[0.25cm]
$\nu_\mu\; (n/p)\rightarrow\nu_\mu\pi^0\;(n/p)$ (NC $\pi^0$)    & $1.2$ \\[0.25cm]
$\nu_\mu\; (p/n)\rightarrow\nu_\mu\pi^\pm\; (n/p)$ (NC $\pi^\pm$) & $0.7$ \\[0.25cm]
Other        & $0.9$ \\[0.1cm] \hline \hline
\end{tabular}
\caption{Yearly rate of $\nu_\mu$ event types in a MiniBooNE-like detector at 
an 8 GeV, 2 MW proton driver. (Data from \cite{MBRunPlan:2003}).}
\label{tab:scaled_mb}
\end{center}
\end{table}

MiniBooNE's design proton intensity calls for $5 \EE{12}$ protons per spill, 
repeating at 5 Hz. For a $10^7$-s operational year this corresponds to 
$2.5 \EE{20}$ P/yr. An 8 GeV, 2 MW proton driver would deliver 
$1.6 \EE{22}$ P/yr.  Therefore, the event rates with a 2 MW proton driver 
can be obtained by scaling up the MiniBooNE event rates by a factor of 
$160 / 2.5 = 64$. The rates for the dominant neutrino interaction channels 
are shown in Table \ref{tab:scaled_mb}.

\end{section}

\begin{section}{Decay-at-Rest Neutrino Beams}
\label{sect:DAR}
The principal advantage of a neutrino beam from muon and pion decay at rest over beams 
produced in other ways is the extremely well known neutrino energy spectrum. The main 
disadvantage is that the flux is isotropic and drops as 1/R$^2$ compared to the high 
degree of collimation of other beam types.

\begin{subsection}{The Beam}
A multi-GeV proton beam will interact in a target to produce pions 
and kaons. If the target is sufficiently large the charged pions will 
lose energy and come to rest. As they slow down some of 
the pions will decay to produce daughter muons, which will also lose 
energy and come to rest.  The fate of the charged pions and muons will 
depend upon whether they are positively or negatively charged. 
The stopped positively charged pions and muons will decay at rest, 
producing a ``prompt'' $\num$ flux from $\pip$ decay followed by 
a ``delayed'' $\nue$ and $\anum$ flux from the much slower 
$\mup$ decays. This time structure will be evident if the incoming 
primary proton beam bunch is short compared to the muon lifetime.
In contrast, before they can decay most of the stopped negatively 
charged pions and muons will be absorbed by the atomic nuclei  
(the relevant capture rates are much larger than the decay rates).
The suppression of negatively charged pion decays is enhanced in 
high-$Z$ target materials for which the capture time is very short. 

Pion and muon decays at rest (DAR) produce an isotropic flux of low energy 
$\nue$, $\num$, and $\anum$ with energies between a few MeV and half 
the meson rest-mass energy.  Neutrino spectra from DAR are shown in 
Fig.~\ref{fig:nuspect-DAR}. Since the flux is isotropic, the challenge is 
to obtain sufficiently high intensities to be of interest to cutting-edge 
experiments. The key to meeting this challenge is to start with a very high 
intensity proton source. For this reason there has been some interest 
\cite{VanDalen:2003fg} in the possibility of exploiting the soon-to-be-completed 
Oak Ridge SNS facility as a low energy neutrino source. The SNS will provide 
a 1.4 MW primary proton beam at 1.3 GeV with a pulse length of 695 ns. The 
pulse length is short compared to the 2~$\mu$s muon lifetime, and hence 
suitable time cuts can separate the neutrino flux produced from the decaying 
pions from the flux produced by decaying muons.  During the duration of the 
proton pulse (from $t = 0$ ns to $t = 695$ ns) 96\% of the positive pions decay. 
From $t = 695$ ns to $t = 5$ $\mu$s the remaining 4\% of the positive pions 
decay and 74\% of the muons decay.

\begin{figure}
\begin{picture}(450,250)
\put(45, 15){\input{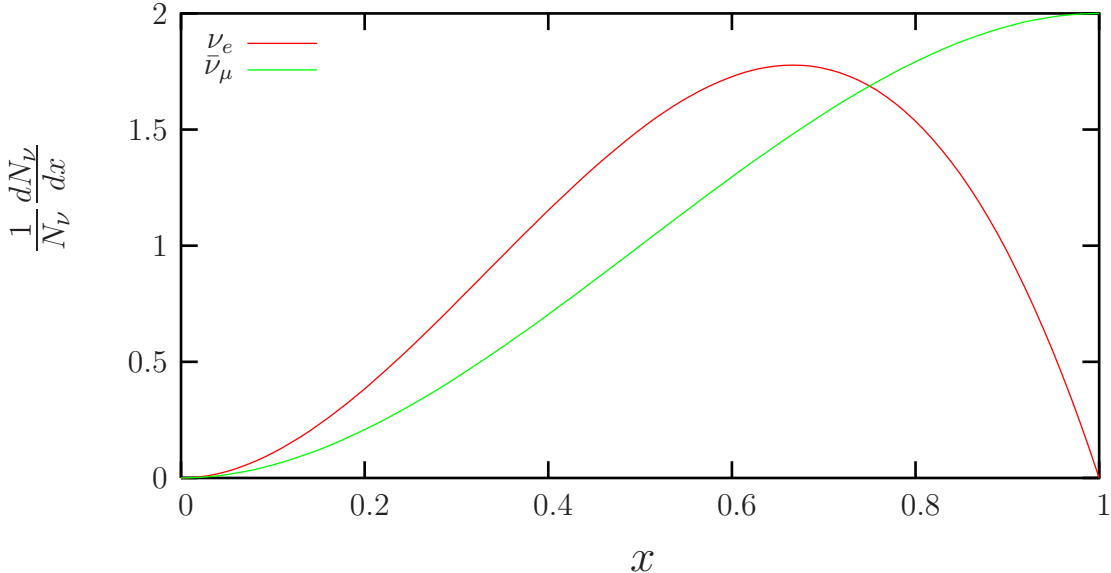}}
\put(15,120){\rotatebox{90}{\Large $\frac{1}{N_{\nu}} \frac{dN_{\nu}}{dx}$}}
\put(250, 0){\Large $x$}
\end{picture}
\caption{The normalized energy spectra of the neutrinos produced from 
$\mu^{+}$ decay at rest, where $x = \frac{2E_{\nu}}{m_{\mu}}$ is the 
fraction of the maximum available energy given to the neutrino during decay.}
\label{fig:nuspect-DAR}
\end{figure}

To a first approximation, at fixed beam power, low energy pion production 
rates are independent of proton beam energy. Hence a first guess at the 
neutrino fluxes from a DAR source at a 2~MW Fermilab proton driver can be 
based on DAR studies for the SNS, scaling upwards by a factor of about 1.4. 
To obtain a better estimate we have used the MARS code and simulated 
particle production, interaction, energy loss, and decay for an 8 GeV 
primary proton beam incident on a carbon target with dimensions sufficiently 
large to contain all of the produced pions. For comparison we have also 
simulated a 5 GeV proton beam incident on the target. To calculate neutrino 
event rates we have used the calculated SNS DAR rates and scaled these 
numbers to correspond to our calculated fluxes, which assume a 2~MW beam at 
8~GeV (or 1.25 MW at 5 GeV), and a MiniBooNE-like detector placed 90 degrees 
off proton-beam axis at a distance of 60 m from the target.  The resulting 
event rates are shown in Table~\ref{tab:DAR} for an operational year 
($10^7$~s).  Note that the 2~MW, 8~GeV proton source produces DAR neutrino 
fluxes that are a factor of 2.5 higher than a potential SNS DAR source, 
somewhat better than suggested by our naive estimate.

\begin{table}[t]
\begin{center}
\begin{tabular}{|r|c|c|c||} \hline
                          &               &               &               \\[-0.5cm]
                          &  FNAL (8 GeV) &  FNAL (5 GeV) &  SNS          \\ 
\hline                    &               &               &               \\[-0.25cm]
$P$/yr                    & $1.6 \EE{22}$ & $1.6 \EE{22}$ & $6.7 \EE{22}$ \\[0.25cm]
\hline                    &               &               &               \\[-0.25cm]
DAR $\nu$       ($\nu/P$) & $1.5$         & $0.9$         & $0.13$        \\[0.1cm]
               ($\nu$/yr) & $2.3 \EE{22}$ & $1.4 \EE{22}$ & $0.92 \EE{22}$\\[0.25cm] 
\hline                    &               &               &               \\[-0.25cm]
$\nu + $e       Events/yr &      2215 \er &      1347 \er &       886 \er \\[0.25cm]
$\nue + $C (CC) Events/yr &     19825 \er &     12054 \er &      7930 \er \\[0.25cm]
$\nu + $C (NC)  Events/yr &     24133 \er &     14673 \er &      9653 \er \\[0.25cm] 
\hline \hline
\end{tabular}
\caption{Neutrino production rates from a carbon target and the event rates in a 
MiniBooNE-like detector placed 90 degrees off axis, 60 m from the target.  The 
production rates are computed using MARS with an 8 GeV, 2 MW proton driver for a 
$10^7$-s operational year.  The event rates in the detector are found by 
scaling from G. VanDalen's paper \cite{VanDalen:2003fg} with fluxes computed by 
MARS for an 8 GeV, 2 MW proton driver in operation for a $10^7$-s year.  Dependence 
of the rates on the neutrino reconstruction efficiency, \er, is explicitly indicated.
The symbol $\nu$ indicates all neutrino and antineutrino flavors.}
\label{tab:DAR}
\end{center}
\end{table}

\end{subsection}

\begin{subsection}{The Physics}
There are a number of fundamental physics measurements one can consider with a decay 
at rest beam. If MiniBooNE should confirm the indication of neutrino oscillations 
reported by the LSND experiment then a new generation of experiments investigating 
the oscilllation in fine detail and, in particular, its energy shape will be needed. 
Besides having an extremely well characterised neutrino spectrum the backgrounds in 
an LNSD- or MiniBooNE-style detector would be extremely low from a proton driver DAR 
beam (the beam related backgrounds are extremely low and a small machine duty factor 
would ensure very low cosmic-ray related backgrounds). This arrangement might 
be the ideal followup to MiniBooNE if that experiment should see an oscillation signal. 
Using the rates from Table~\ref{tab:DAR} a MiniBooNE sized detector operating at an 8 
GeV proton driver would see many hundreds of oscillations events per year (varying with 
the exact oscillation parameters) with about 5 beam unrelated background events and 
a couple of beam related backgrounds (including muon misIDs).  These numbers are 
scaled from \cite{VanDalen:2003fg}.  The low backgrounds and well-known spectrum at 
a DAR source more than make up for the drop in neutrino cross-section at the resulting 
comparatively low neutrino energies.

It has been suggested \cite{Bulanov:2004} that a neutrino beam from muon DAR
can be used to search for a non-zero $\theta_{13}$, the only unmeasured angle 
in the three neutrino mixing matrix. The effect of $\theta_{13}$ on $\nue$ appearance 
is maximised when the oscillation baseline is at the atmospheric $\Delta m^2$. For 
muon DAR energies this puts the detector about 10km from the source and a 
very massive detector (on the scale of the proposed HyperK) would be required. Whilst 
this may be possible it is unlikely to happen on a timescale competitive
with reactor or superbeam experiments.

Table~\ref{tab:DAR} shows a healthy rate of neutrino electron scattering in a 
MiniBooNE scale detector at an upgraded FNAL proton driver. These events can be 
used to make a very low Q$^2$ measurement of $\sin^2\theta_w$, the Weak mixing 
angle, by measuring the cross-section ratio
\begin{equation}
R = \frac{\sigma(\num e)}{\sigma(\nue e) + \sigma(\anum e)} \;\; .
\end{equation}
Neutrino electron scattering events can be identified by looking for very forward 
going electrons in the detector, and events from the numerator of the $R$ ratio 
come in time with the beam whereas events in the denominator come later (in 
general).  Assuming all neutrinos have the same flux shape, and neglecting terms 
of order $m_e/E_\nu$, this ratio calculated in the SM is
\begin{equation}
R = \frac{0.75 - 3\sin^2\theta_w + 4\sin^4\theta_w}{1 + 
    2\sin^2\theta_w + 8\sin^4\theta_w} \;\; .
\end{equation}
Based on the work of \cite{Imlay:2003} a measurement of $R$ in this way at an 
upgraded FNAL proton driver would yield a 1-2\% measurement of $\sin^2\theta_w$. 
This measurement would be at low Q$^2$ compared to other weak mixing angle 
measurements. In fact the only other measurement in this Q$^2$ range comes from 
the totally different regime of atomic parity violation. In view of the recent 
indications of non-SM behaviour in the extraction of $\sin^2\theta_w$ from 
neutrino electron scattering at higher Q$^2$ \cite{NuTeV2002} this measurement 
from a decay at rest beam from an upgraded FNAL proton driver is very attractive.

Neutrino electron scattering can also be used to search for a neutrino magnetic 
moment as it adds an electromagnetic component to the otherwise weak neutrino 
electron scattering cross-section. This component gets stronger with decreasing 
electron recoil energy and so low energy neutrino sources such as reactors or 
stopped muon decay are the best places to go looking for it. The well known energy 
spectrum of decay at rest neutrinos is a particular advantage when looking for a 
distortion in the recoil electron energy spectrum caused by a neutrino magnetic 
moment.

\end{subsection}

\end{section}

\begin{section}{Neutrino Beam from a Solenoid Decay Channel}
\label{sect:DIF}

\begin{subsection}{The Beam}
Neutrino beams are usually produced using a magnetic horn to sign select and 
focus the secondary charged pions into an approximately parallel beam 
that then propagates within a long large aperture decay pipe. 
An alternative approach is to use the optics that has been 
proposed for the upstream stage of a neutrino factory, in which 
a pion production target is within a high-field solenoid with a field 
strength $B_0$, sufficiently high to radially contain essentially  
all of the produced charged pions. This means that, if $p_T$ is the 
maximum transverse momentum of the produced pions, then $B_0$ must be 
chosen such that 
\begin{equation}
p_T < e \, B_0 \, \left( \frac{R_0}{2} \right) \;\; ,
\end{equation}
where $R_0$ is the radius of the beampipe around the target. 
Note that for an 8~GeV primary proton beam incident on a carbon target 
the secondary pion transverse momentum peaks at about 150~MeV/c. 
For the design we have explored $R_0 = 7.5$ cm and $B_0 = 20$ T. This 
system will capture particles with $p_T < 225$ MeV/c, resulting in a 
captured beam with significantly broader energy spectrum than one captured 
with a horn.  

It would be unnecessarily expensive to extend the 20~T solenoid for the full 
length of the decay channel downstream of the target.  Therefore, downstream 
of the target the charged pions, and their daughter muons, are confined radially 
within a lower field ($B = 1.25$ T) larger radius ($R = 30$ cm) solenoid 
decay channel.  To ensure efficient transfer of the beam from the 
high-field to the low-field solenoids, the distance over which the field is 
decreased must be approximately 10 m or more.  Within this section the focusing
field strength $B(s)$ is adiabatically decreased whilst the beampipe radius 
$R(s)$ is increased, keeping the magnetic flux within the beampipe, 
$\Phi \propto B(s) R(s)^2$, constant.  The continual transverse focusing of the 
beam throughout the entire length of the decay channel allows for the marked 
increase in muon yields over those from horn designs.  A typical layout is 
shown in Figure~\ref{fig:layout}.

\begin{figure}
\begin{picture}(450,65)
\linethickness{2pt}
\put(  0,  0){\includegraphics[width=6.5in]{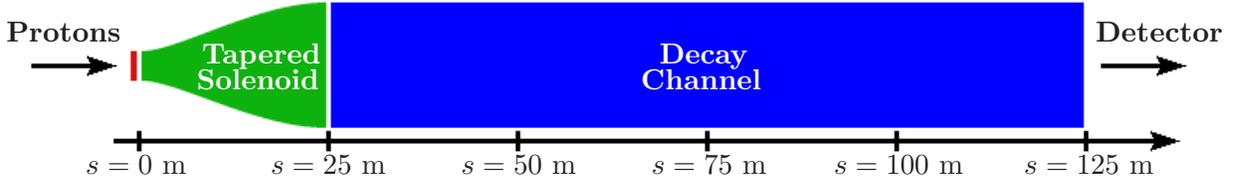}}
\put( 40,  0){\small $s = 0$ m}
\put(110,  0){\small $s = 25$ m}
\put(182,  0){\small $s = 50$ m}
\put(254,  0){\small $s = 75$ m}
\put(323,  0){\small $s = 100$ m}
\put(395,  0){\small $s = 125$ m}
\put( 10, 50){\small \bf Protons}
\put(422, 50){\small \bf Detector}
\put( 84, 42){\small \bf \color{white} Tapered}
\put( 82, 32){\small \bf \color{white} Solenoid}
\put(257, 42){\small \bf \color{white} Decay}
\put(250, 32){\small \bf \color{white} Channel}
\end{picture}
\caption{This figure depicts the typical layout of a solenoid-based pion 
capture and decay channel used for generating neutrino beams.}
\label{fig:layout}
\end{figure}

The charged pions within the solenoid channel decay via 
$\pip \to \mup \, \num$ and $\pim \to \mum \, \anum$, with a lifetime of 
$\tau_{\pi} = 26$ ns.  The daughter muons decay with a lifetime of 
$\tau_{\mu} = 2.2$ $\mu$s via  $\mup \to \elp \, \nue \, \anum$ and 
$\mum \to \elm \, \anue \, \num$.  Due to the significant difference between 
the pion and muon lifetimes, the muon content of the beam will steadily 
increase over the first $\sim$75 m downstream from the target.  At greater 
distances the muon content of the beam decreases as the muons decay.  The 
muon and charged pion content of the beam is shown in Table~\ref{tab:charged} 
as a function of distance downstream from the target, $s$, for an 8 GeV, 2 MW 
primary proton beam incident on a carbon target with a 25 m adiabatic section 
following the 20 T capture solenoid (Figure \ref{fig:layout}).  The corresponding 
evolution of the charged pion and muon spectra is shown in Figure 
\ref{fig:cpSpec}.  Note that the simulated layout does not sign select the pion 
and/or muon beam.  Hence, downstream of the decay channel, the neutrino beam 
contains neutrinos from $\pi^+, \pi^-, \mu^+$ and $\mu^-$ decays. However, the 
design can be modified by placing suitable large angle bends at any desired 
positions along the decay channel.  Thus sign selection bends, and bends to 
enhance of the muon DAR component, are possible options.

\begin{table}[b]
\begin{center}
\begin{tabular}{|l|c|c|c|c|c||} \hline
         &          &          &          &           &          \\[-0.5cm]
         & $s=25$ m & $s=50$ m & $s=75$ m & $s=100$ m & $s=125$ m\\[0cm]
\hline
         &          &          &          &           &          \\[-0.25cm]
$\mup/P$ & 0.16     & 0.20     & 0.21     & 0.21      & 0.22     \\[0.25cm]
$\mum/P$ & 0.16     & 0.20     & 0.21     & 0.21      & 0.21     \\[0.25cm] 
\hline
         &          &          &          &           &          \\[-0.25cm]
$\pip/P$ & 0.095    & 0.051    & 0.030    & 0.020     & 0.014    \\[0.25cm]
$\pim/P$ & 0.087    & 0.044    & 0.025    & 0.016     & 0.011    \\[0.25cm] 
\hline \hline
\end{tabular}
\caption{The number of charged particles per incident proton in the beam at 
various distances downstream from the target.  These numbers are computed by 
MARS for an 8 GeV, 2 MW proton driver such as the proposed Fermilab upgrade.}
\label{tab:charged}
\end{center}
\end{table}

\begin{figure*}
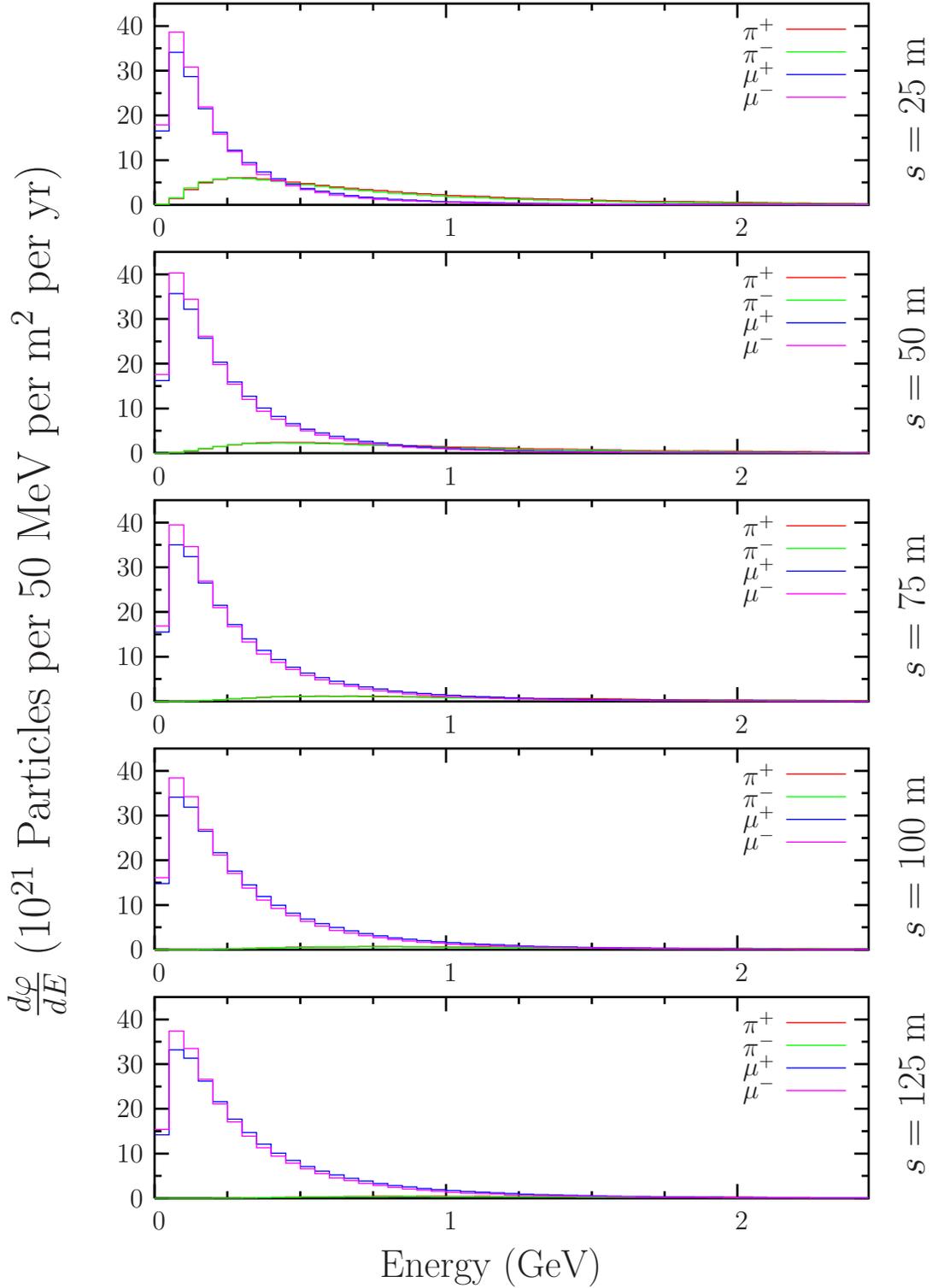

\begin{center}
\begin{picture}(450,580)
\put(50,575){\small Charged Particle Fluxes in the Decay Channel with 
$1.6\times10^{22}$ POT at 8 GeV}
\put(50,455){\input{cpSpec-0}}
\put(50,345){\input{cpSpec-1}}
\put(50,235){\input{cpSpec-2}}
\put(50,125){\input{cpSpec-3}}
\put(50, 15){\input{cpSpec-4}}
\put(15,115){\rotatebox{90}{\LARGE $\frac{d\varphi}{dE}$ ($10^{21}$ Particles 
per 50 MeV per m$^2$ per yr)}}
\put(180, 0){\Large Energy (GeV)}
\put(410,485){\rotatebox{90}{\Large $s = 25$ m}}
\put(410,375){\rotatebox{90}{\Large $s = 50$ m}}
\put(410,265){\rotatebox{90}{\Large $s = 75$ m}}
\put(410,150){\rotatebox{90}{\Large $s = 100$ m}}
\put(410, 45){\rotatebox{90}{\Large $s = 125$ m}}
\end{picture}
\caption{Charged particle fluxes at various positions down the 100 m decay 
channel with a 30 cm beampipe radius.  We assume an 8 GeV, 2 MW proton driver in 
operation for a $10^{7}$-s year.}
\label{fig:cpSpec}
\end{center}
\end{figure*}

\begin{figure*}
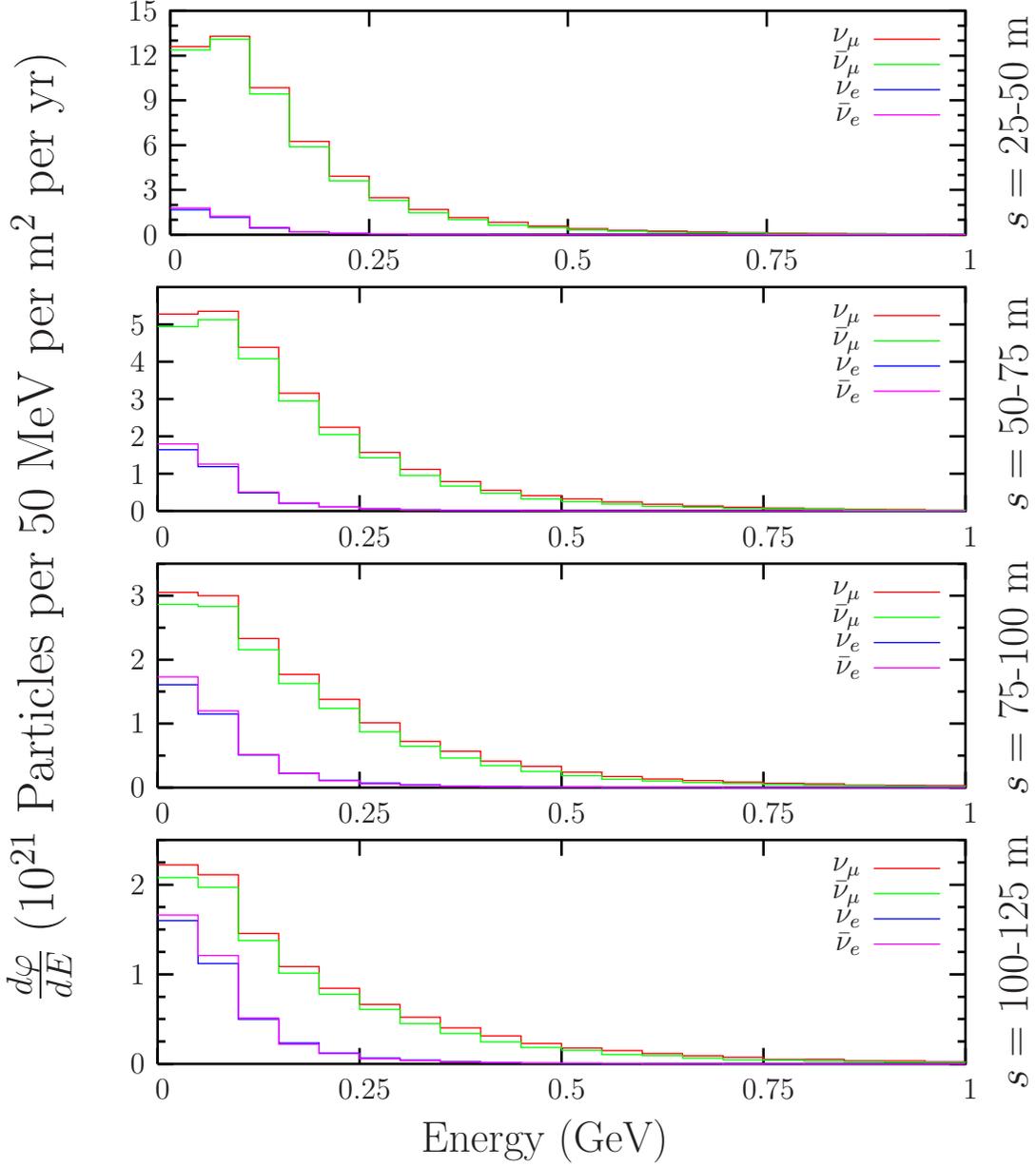

\begin{center}
\begin{picture}(450,470)
\put(50,468){\small Neutrino Fluxes in the Decay Channel with
$1.6\times10^{22}$ POT at 8 GeV}
\put(50,345){\input{nuSpec-1}}
\put(50,235){\input{nuSpec-2}}
\put(50,125){\input{nuSpec-3}}
\put(50, 15){\input{nuSpec-4}}
\put(15, 60){\rotatebox{90}{\LARGE $\frac{d\varphi}{dE}$ ($10^{21}$ Particles
per 50 MeV per m$^2$ per yr)}}
\put(180, 0){\Large Energy (GeV)}
\put(410,367){\rotatebox{90}{\Large $s = 25$-50 m}}
\put(410,255){\rotatebox{90}{\Large $s = 50$-75 m}}
\put(410,142){\rotatebox{90}{\Large $s = 75$-100 m}}
\put(410, 25){\rotatebox{90}{\Large $s = 100$-125 m}}
\end{picture}
\caption{Neutrino fluxes at their point of origin in the four 25m segments down 
the 100 m decay channel with a 30 cm beampipe radius.  We assume an 8 GeV, 2 MW 
proton driver in operation for a $10^{7}$-s year.}
\label{fig:nuSpec}
\end{center}
\end{figure*}


\begin{table}[t]
\begin{center}
\begin{tabular}{|lr|c|c|c|c||} \hline
          &&           &           &            &             \\[-0.5cm]
          && $25-50$ m & $50-75$ m & $75-100$ m & $100-125$ m \\[0cm] 
\hline
          &&           &           &            &             \\[-0.25cm]
$\num/P$  & {\small \it produced} &
  0.047     & 0.023     & 0.014      & 0.0095      \\
          & {\small \it reaching detector} & 
  0.0034    & 0.0026    & 0.0023     & 0.0022      \\[0.25cm]
$\anum/P$ & {\small \it produced} & 
  0.046     & 0.021     & 0.012      & 0.0086      \\ 
          & {\small \it reaching detector} & 
  0.0028    & 0.0021    & 0.0019     & 0.0017      \\[0.25cm] 
\hline
          &           &           &            &             \\[-0.25cm]
$\nue/P$  & {\small \it produced} &
  0.0031    & 0.0034    & 0.0033     & 0.0034      \\
          & {\small \it reaching detector} &
  0.000050  & 0.00010   & 0.00014    & 0.00016     \\[0.25cm]
$\anue/P$ & {\small \it produced} &
  0.0032    & 0.0035    & 0.0034     & 0.0034      \\
          & {\small \it reaching detector} &
  0.000046  & 0.000091  & 0.00013    & 0.00015     \\[0.25cm]
\hline \hline
\end{tabular}
\caption{The number of neutrinos produced per incident 
proton in the four 25 m segments of the 100 m decay channel, as well as those
neutrinos produced that reach the detector.  These numbers are computed by 
MARS for an 8 GeV, 2 MW proton driver such as the proposed Fermilab upgrade.}
\label{tab:nuProd}
\end{center}
\end{table}

The flavor content of the neutrino beam depends upon the length of the 
decay channel (Table \ref{tab:nuProd} and Figure \ref{fig:nuSpec}). Early in 
the channel pion decays dominate. In the last 25 m segment of the decay 
channel the pions constitute a few percent of the charged particle beam, and 
the number of muon and pion decays are similar.  Figures \ref{fig:nupSpec-X} and 
\ref{fig:numSpec-X} show the neutrino fluxes from neutrinos born in the last 50 m 
of the decay channel at a distance of 100 m from the end of the decay channel.  
The fluxes are averaged over a 10 m radius from the axis of the beam.  Figure 
\ref{fig:nupSpec-X} shows the neutrino fluxes produced solely from pion decays, 
and Figure \ref{fig:numSpec-X} shows the corresponding neutrino fluxes from muon 
decays.  It is worth noting that, while the number of pion and muon decays in 
the last 50 m of decay channel are comparable, the neutrinos coming from pion 
decays are considerably more collimated because of their comparatively higher 
energy in the last 50 m of the decay channel (Figures \ref{fig:nuSpec} and 
\ref{fig:nuDist-X}).

\begin{figure}
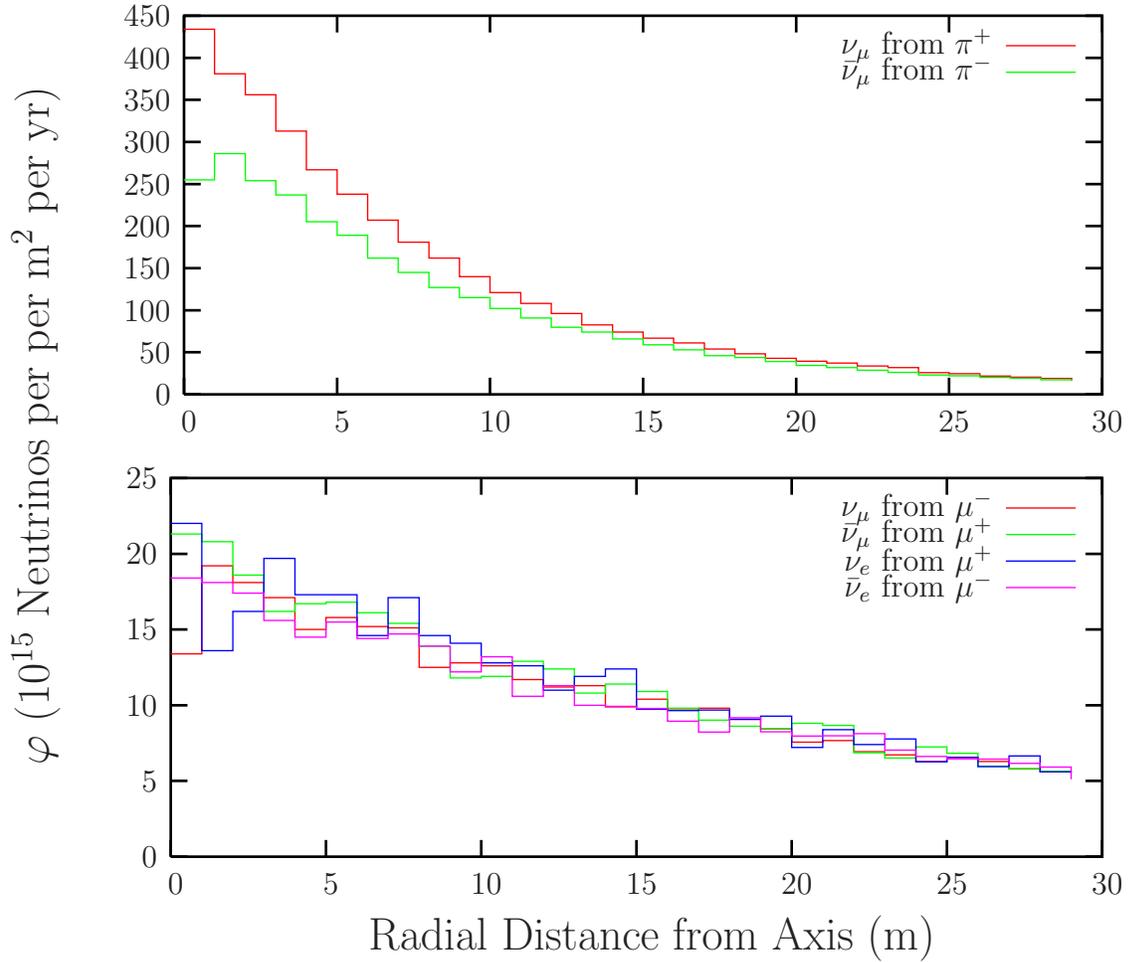

\begin{picture}(450,380)
\put(60,370){\small Neutrino Radial Distribution at the Detector with 
$1.6\times10^{22}$ POT at 8 GeV}
\put(45,190){\input{nupDist-X}}
\put(45, 15){\input{numDist-X}}
\put(15, 70){\rotatebox{90}{\Large $\varphi$ ($10^{15}$ Neutrinos per 
per m$^2$ per yr)}}
\put(150, 0){\Large Radial Distance from Axis (m)}
\end{picture}
\caption{This figure shows the fluxes, at a distance of 100 m from the end of 
the solenoid decay channel and averaged over 1m-radius annuli, of the neutrinos 
produced in the last 50 m of decay channel.  We assume an 8 GeV, 2 MW proton 
driver with operation over a $10^7$-s year.}
\label{fig:nuDist-X}
\end{figure}

\begin{figure}
\begin{picture}(450,380)
\put(60,370){\small Neutrino Fluxes from $\pi^{\pm}$ Decays at the Detector with 
$1.6\times10^{22}$ POT at 8 GeV}
\put(50,190){\input{numupSpec-X}}
\put(50, 15){\input{anumupSpec-X}}
\put(15, 32){\rotatebox{90}{\Large $\frac{d\varphi}{dE}$ ($10^{15}$ Neutrinos per 
50 MeV per m$^2$ per yr)}}
\put(220, 0){\Large Energy (GeV)}
\end{picture}
\caption{This figure shows the fluxes, at a distance of 100 m from the end of 
the solenoid decay channel and averaged over a radius of 10 m, of the neutrinos 
produced solely from pion decay in the last 50 m of decay channel.  We assume 
an 8 GeV, 2 MW proton driver with operation over a $10^7$-s year.}
\label{fig:nupSpec-X}
\end{figure}

\begin{figure}
\begin{picture}(450,380)
\put( 75,370){\small Neutrino Fluxes from $\mu^{\pm}$ Decays at the Detector with 
$1.6\times10^{22}$ POT at 8 GeV}
\put( 50,190){\input{numumSpec-X}}
\put(250,190){\input{nuelSpec-X}}
\put( 50, 15){\input{anumumSpec-X}}
\put(250, 15){\input{anuelSpec-X}}
\put( 15, 32){\rotatebox{90}{\Large $\frac{d\varphi}{dE}$ ($10^{15}$ Neutrinos 
per 50 MeV per m$^2$ per yr)}}
\put(220,  0){\Large Energy (GeV)}
\end{picture}
\caption{This figure shows the fluxes, at a distance of 100 m from the end of 
the solenoid decay channel and averaged over a radius of 10 m, of the neutrinos 
produced solely from muon decay in the last 50 m of decay channel.  We assume 
an 8 GeV, 2 MW proton driver with operation over a $10^7$-s year.}
\label{fig:numSpec-X}
\end{figure}

To calculate event rates we consider a MiniBooNE-like detector placed on 
axis at a distance of 100 m from the end of the decay channel.  
The rates for different interactions and neutrino types, assuming an 8 GeV, 2 MW 
proton driver in operation for a $10^7$-s year, are shown in Table 
\ref{tab:nuRate}. The cross-sections on CH$_2$ are computed using the NUANCE code 
\cite{nuance}.

\begin{table}
\renewcommand{\arraystretch}{1.5}
\begin{center}
\begin{tabular}{|r|c|c|c|c|c|c|} \hline
&  \multicolumn{6}{|c|}{Million events per year}  \\ \cline{2-7}
\multicolumn{1}{|c|}{$\nu$ event type}          
&  $ \num$ ($\pip$) & $\anum$ ($\pim$) & $ \num$ ($\mum$) & $\anue$ ($\mum$) & $\anum$ ($\mup$) & $ \nue$ ($\mup$)  \\ \hline
$\nu (n/p)\rightarrow l^\pm(p/n)$ (CC QE)                   & 24.9  & 6.5  & 0.83  & 0.37  & 0.95  & 0.28 \\
$\nu\; (n/p)\rightarrow\nu\; (n/p)$ (NC QE)                 & 10.8  & 3.1  & 0.39  & 0.18  & 0.40  & 0.13 \\
$\nu\; (n/p)\rightarrow l^\pm\pi^\mp\;(n/p)$ (CC $\pi^\pm$) &  5.2  & 0.5  & 0.12  & 0.03  & 0.16  & 0.02 \\
$\nu (n/p)\rightarrow l^\pm\pi^0 (p/n)$ (CC $\pi^0$)        &  0.9  & 0.2  & 0.02  & 0.01  & 0.03  & 0.00 \\
$\nu\; (n/p)\rightarrow\nu\pi^0\;(n/p)$ (NC $\pi^0$)        &  1.5  & 0.4  & 0.04  & 0.02  & 0.04  & 0.01 \\
$\nu\; (p/n)\rightarrow\nu\pi^\pm\; (n/p)$ (NC $\pi^\pm$)   &  0.9  & 0.2  & 0.02  & 0.01  & 0.02  & 0.01 \\
Other                                                       &  1.8  & 1.3  & 0.05  & 0.07  & 0.06  & 0.04 \\ \hline
\end{tabular}
\caption{The above table shows the total integrate events rates from neutrinos 
produced in the last 50 m of the decay channel from pion and muon decays.  The 
MiniBooNE-like detector is placed at 100 m from the end of the decay channel. 
The fluxes are computed by MARS for an 8 GeV, 2 MW proton driver in operation 
for a $10^7$-s year and the cross-sections on CH$_2$ are computed by the NUANCE 
code \cite{nuance}.}
\label{tab:nuRate}
\end{center}
\end{table}






\end{subsection}

\begin{subsection}{The Physics}
As can be seen in Figure~\ref{fig:nupSpec-X}, the $\nu_\mu$ ($\bar{\nu}_\mu$) 
flux from $\pi^{+}$ ($\pi^{-}$) decays-in-flight resulting from this neutrino
beam are large.  The resulting neutrino event rates in a 
MiniBooNE-like detector 100m from the end of the decay channel will
be unprecedented and could be used to precisely measure the neutrino
oscillation parameters for both $\nu$ and $\bar{\nu}$ in the 
$L/E \approx 1$ range as is currently being investigated by the 
MiniBooNE experiment[3]. 

If MiniBooNE confirms the LSND result, this will indicate
neutrino oscillations in the $L/E \approx 1$,  
$\Delta m^2 \approx 1$eV$^2$ 
range, and a systematic study of this
effect, for both $\nu$ and $\bar{\nu}$, will be required.  A
solenoid-based neutrino beam with a large angle bend to sign select
the pions would be an excellent source for these studies.  The 
bend will allow for a relatively pure $\nu$ or $\bar{\nu}$ beam, 
depending on the polarity of the bending magnet.   The polarity reversal
could take place on the timescale of hours or minutes, thus reducing
systematic uncertainties between $\nu$ and $\bar{\nu}$ running
arising from detector response.

For these experiments, two large ($\approx 1$kton) detectors, would
be required at distances on the order of 100m and 250m (the exact
locations would be determined by the neutrino data at the time of 
experiment construction).  In addition, a relatively
small detector ($\approx 20$ tons) would be situated at a near 
location ($\approx 25-50$m).  This detector would be highly segmented
to precisely monitor the flux $\times$ cross section and would
have charge identification to measure the ``wrong-sign'' component
of the neutrino beam.

With this configuration of neutrino beam and detectors, 
simultaneous precision measurements of $\nu_\mu \rightarrow \nu_e$, 
$\bar{\nu}_\mu \rightarrow \bar{\nu}_e$, $\nu_\mu \rightarrow \nu_x$,
and $\bar{\nu}_\mu \rightarrow \bar{\nu}_x$ in the $\Delta{m}^2=$.1-1 
eV$^2$ region would be achievable.  With the correct arrangement of 
detectors, an oscillation measurement using neutrinos from $\mu^+$ that 
decay-at-rest in the beam stop could also be conducted in parallel.  
These neutrinos have a different time structure due to the 
long-lifetime of the $\mu^+$ and are of much lower energy as described 
above.  These factors will allow for a clean separation of the neutrino 
sources.

In addition to servicing the oscillation measurements, the near detector 
would be a rich source of non-oscillation neutrino physics.  A 20 ton
detector at 25m would collect events at 10\% of the rate shown in
Table 5.  This would allow measurements of neutral- and 
charged-current processes with unprecedented precision.  These details
of these processes (cross-sections, energy- and $Q^2$-dependence)
are needed for oscillation measurements and provide valuable information
on nucleon structure.   

\end{subsection}

\end{section}

\begin{section}{Conclusions}
\label{sect:concl}
We have considered three options for generating a low energy neutrino
beam using an upgraded 2 MW Fermilab Proton Source; (i) a MiniBooNE-like
beam using a conventional neutrino horn, (ii) a decay at rest neutrino
source, and (iii) a neutrino beam downstream of a solenoid decay channel.
In all three cases the fluxes are sufficiently large to provide an
interesting physics program. The upgraded MiniBooNE-like beam would be a
factor of 64 more intense than the existing beam. The decay at rest beam
would offer an intensity that is a factor of 2.5 higher than the
corresponding intensity at an equivalent source at the SNS. The solenoid
decay channel beam has some characteristics which would make it
particularly interesting if MiniBooNE were to confirm the LSND result,
provided sign-selection can be implemented efficiently in a (bent)
solenoid.
\end{section}


\end{document}